\documentclass[11pt]{article}
\usepackage{algorithm}

\usepackage[final]{acl}

\usepackage{amsfonts}
\usepackage{amsmath}
\usepackage{amssymb}
\usepackage{amsthm}
\usepackage{dsfont}
\usepackage{times}
\usepackage{latexsym}
\usepackage{subcaption}

\usepackage{algorithm}
\usepackage{algpseudocode}
\usepackage{booktabs} 

\usepackage[T1]{fontenc}

\usepackage[utf8]{inputenc}

\newtheorem{theorem}{Theorem}
\newtheorem{definition}{Definition}
\newtheorem{remark}{Remark}
\usepackage{microtype}

\usepackage{inconsolata}

\usepackage{graphicx}

\newcommand{\eps}{\varepsilon}
\title{Can We Share Speech Data With Differential Privacy?}

\author{Efthymios Tsaprazlis$^*$\, Vasilis Varsamis$^*$ \, Tiantian Feng\footnotemark[2]  \\
\textbf{Sai Praneeth Karimireddy \, Shrikanth Narayanan}  \\
University of Southern California\\
Los Angeles, CA 90089, USA \\
\texttt{\{tsaprazl,varsamis\}@usc.edu}
}

\begin{document}
\begingroup
  \renewcommand{\thefootnote}{\fnsymbol{footnote}}
  \maketitle
  \footnotetext[2]{Now at CUHK-Shenzhen, work done while at USC}
  \def\thefootnote{*}\footnotetext{Equal contribution}
\endgroup
\setcounter{footnote}{0}

\begin{abstract}

We study whether speech can be shared with formal privacy guarantees. We present speaker anonymization mechanisms with local differential privacy, acting on the speaker embedding of a zero-shot TTS system: one adds calibrated Laplace noise, the other selects a surrogate identity from a public donor pool via the exponential mechanism. We find that additive noise requires prohibitively large privacy budgets, whereas selection reaches the same empirical privacy at far smaller budgets. Our analysis identifies practical operating points for differential privacy in speech, reveals worst-case acoustic identity leakage, and quantifies the impact of informed adversaries.

\end{abstract}

\section{Introduction}

Speech recognition and synthesis systems have become central to how people interact with devices, share information, and communicate. Yet spoken data is uniquely sensitive, as a single utterance simultaneously reveals what was said and who said it. Regulatory frameworks such as GDPR~\citep{gdpr2016} explicitly classify voice as personal data and, in some cases, as biometric data, reflecting the fact that a speaker's voiceprint is as identifying as a fingerprint. As speech-enabled systems grow more capable and pervasive, the need for rigorous, verifiable privacy guarantees over spoken data becomes increasingly urgent.

The dominant approach to mitigating speaker identity leakage is \emph{speaker anonymization}: transforming an utterance to conceal the speaker's identity while preserving its linguistic content~\citep{EURECOM+7703}. Modern anonymization systems achieve this by decomposing speech into an acoustic channel, which carries speaker-specific information such as timbre and prosody, and a linguistic channel, which carries the spoken content. They then privatize the acoustic channel to suppress speaker-identifying information and resynthesize the speech by combining this perturbed acoustic channel with the intact linguistic content. Despite their empirical success, these systems offer no worst-case guarantees.  We aim to close this gap by providing formal local differential privacy (DP) guarantees on the released speech. Our first approach injects calibrated Laplace noise into the speaker-conditioning embedding at inference time. The synthesizer depends only on this privatized embedding and requires no retraining. Crucially, because linguistic content can leak into the speaker embedding, allowing the text to vary across utterances can artificially inflate empirical privacy. By holding the spoken text fixed across speakers during evaluation, we remove text-induced variation and obtain a conservative empirical test of acoustic identity leakage. 

However, while additive mechanisms can protect identity, they waste privacy budget perturbing the embedding's magnitude. Because standard speaker-verification adversaries~\citep{EURECOM+7703,desplanques20_interspeech} compare embeddings by cosine similarity, they are invariant to magnitude shifts. Consequently, any budget spent altering the magnitude buys no protection. We therefore propose to \emph{replace} the source embedding instead of perturbing it. That is, we do not release a noisy version of the speaker's own embedding, but the embedding of a different, real speaker. Candidates come from a public donor pool that is disjoint from the speakers being protected. We score each candidate by its cosine similarity to the source embedding and sample one with the exponential mechanism~\citep{mcsherry2007mechanism}, which favors high-scoring candidates to a degree set by the privacy budget $\varepsilon$. A smaller $\varepsilon$ makes the draw closer to uniform over the pool, giving stronger privacy, whereas a larger $\varepsilon$ concentrates it on the most similar donors, giving better utility. Since cosine similarity ignores magnitude, the score depends only on the embedding's direction, which is exactly what verification adversaries compare, so none of the budget is wasted on magnitude. Since the score is also bounded, the selection is $\varepsilon$-locally differentially private. The mechanism can be applied per utterance or per speaker; at the speaker level, every utterance of a speaker maps to the same surrogate.

\paragraph{Our Contributions}
\begin{enumerate}
    \item We provide formal local differential privacy guarantees for speaker anonymization, operating directly on the speaker embedding and requiring no model retraining.
    \item We instantiate two additive mechanisms, coordinate-wise and radial Laplace, and characterize their operating regimes.
    \item We introduce a selection mechanism that replaces the speaker's identity with a surrogate drawn from a public donor pool via the exponential mechanism~\citep{mcsherry2007mechanism}. When applied at the speaker level, it assigns the same surrogate across all of a speaker's utterances while satisfying differential privacy.
    \item We present a practical anonymization pipeline built on these mechanisms and Qwen3-TTS~\citep{hu2026qwen3} that achieves privacy with minimal utility loss.
\end{enumerate}
\section{Related Work}

\paragraph{Speech Privacy}
The speech-privacy domain is organized around standardized benchmarks such as the VoicePrivacy Challenge~\citep{EURECOM+7703}, which supply the datasets and protocols for measuring how well a system conceals speaker identity while preserving linguistic and emotional content. Empirical methods fall into three categories. \emph{Signal-processing} methods such as the randomized McAdams coefficient~\citep{mcadams1984spectral} reshape the spectral envelope and timbre of the waveform directly~\citep{patino21_interspeech}. \emph{Neural voice conversion} methods extract linguistic bottleneck features via ASR and substitute a pseudo-speaker for the original speaker representation before re-synthesis~\citep{EURECOM+5909,9829284}, sometimes adding adversarial objectives (e.g.\ gradient-reversal layers) to strip residual identity from the representation~\citep{noe21_interspeech, DBLP:conf/interspeech/JalalPZOSHLJ23}. \emph{Generative} approaches use zero-shot neural codecs with semantic disentanglement to remove speaker-dependent variation without target-speaker data~\citep{pmlr-v235-ju24b,turavecino25_interspeech}. These systems set the empirical privacy–utility frontier, but they offer no formal guarantee. Their protection is measured against specific attackers, typically through EER, which averages over the population and can conceal the tail speakers most at risk of linkage. Our goal is not the strongest empirical anonymizer. We ask what a formal $\varepsilon$-LDP guarantee costs in utility, so we compare mechanisms that share the same guarantee under identical data, synthesizer, and evaluation protocol.

\paragraph{Differential Privacy}
Several works seek formal DP guarantees for speech. \citet{shamsabadi2023differentially} extract pitch and bottleneck features with autoencoders whose noise layers add Laplace noise, providing local differential privacy (LDP)~\cite{kasiviswanathan2011can, duchi2013local}. Their guarantee, however, holds at the frame level and composing it across the K frames of an utterance yields a budget that grows with utterance length and is uninformative for realistic K. Our mechanisms act once per utterance or once per speaker, so the budget does not depend on utterance length. Their reported EER also stays short of chance, whereas pool selection reaches chance-level EER at small budgets without degrading synthesis. Furthermore, a related line of work protects individual attributes instead of identity.~\citet{chouchane2023differentially} conceal speaker biological sex by adding coordinate-wise Laplace noise to the latent space of an adversarial autoencoder. Extending these attribute protections to distributed environments, \citet{feng22b_interspeech} incorporate user-level DP within federated learning frameworks to defend speech emotion recognition systems against attribute inference attacks.

A natural attempt to relax LDP for continuous data is metric differential privacy ($d_{\mathcal{X}}$-privacy)~\citep{chatzikokolakis2013broadening}, which scales the indistinguishability bound by the distance between inputs and generalizes the geo-indistinguishability model of location privacy~\citep{andres2013geo}. Concretely, the mechanism fixes a privacy \emph{rate} $\varepsilon$ rather than a budget: two inputs at distance $d$ are only $\varepsilon d$-indistinguishable, so protection is strong for nearby inputs and decays linearly as they separate. A concrete budget appears only once a radius is fixed, where inputs within distance $r$ are protected at level $\varepsilon r$. \citet{feyisetan2020privacy} apply this to natural language by adding noise to each word's embedding and decoding the noisy vector to the nearest word in a fixed vocabulary. The analogous construction for speakers, perturbing the speaker embedding and decoding it to the nearest donor in a public pool, is straightforward in our setting. Its guarantee, however, weakens as speakers move apart in embedding space, so speakers with distinctive voices, the tail speakers most exposed to linkage, would receive the weakest protection. We instead target pure $\varepsilon$-LDP, which bounds the distinguishability of every pair of speakers uniformly.

\section{Privacy Definitions}
\label{sec:privacy}

Differential privacy (DP)~\citep{dwork2006differential} is the standard framework for quantifying the privacy of a system that releases sensitive data. DP guarantees are achievable through simple noise-adding mechanisms and are preserved under post-processing, so any further computation on a privatized quantity does not weaken its guarantee. Following \citet{shamsabadi2023differentially}, we work in the local
model~\citep{kasiviswanathan2011can, duchi2013local}, which is a distributed model where each individual perturbs their own data before release, eliminating the need for a central data curator. The formal definition follows.

\begin{definition}[Local Differential Privacy {\normalfont\citep{kasiviswanathan2011can, duchi2013local}}]
\label{def:ldp}
Let $\mathcal{A}$ be a randomized algorithm that takes inputs in a space $\mathcal{X}$ and $\eps > 0$. We say that $\mathcal{A}$ is $\eps$-\emph{LDP} if for every pair of inputs
$\mathbf{x}, \mathbf{x'} \in \mathcal{X}$ and every measurable
$S \subseteq range(\mathcal{A})$,
\[
  \Pr[\mathcal{A}(\mathbf{x}) \in S] \;\le\; e^{\eps}\,\Pr[\mathcal{A}(\mathbf{x'}) \in S].
\]
\end{definition}
Local DP says that no change to the input shifts the output distribution by more than a factor of $e^\eps$, with smaller $\eps$ yielding stronger privacy. In our case, $\mathcal{A}$ perturbs the speaker embedding, so $\varepsilon$ bounds how distinguishable two speakers remain from the released audio. The transcript is preserved, not privatized, so the guarantee concerns speaker identity rather than spoken content.

\paragraph{Worst-Case Privacy and Its Measurement}
Definition~\ref{def:ldp} is a \emph{worst-case} statement: the bound holds for every pair of inputs and therefore every adversary, not for a typical speaker
or an average over the corpus. We hold the empirical evaluation to the same standard, reporting TPR at low FPR rather than EER alone
(Section~\ref{sec:tpr}).

\paragraph{Voice and Content Are Not Fully Independent} Each utterance can be viewed as a pair of information flows $(i, t_i)$ consisting of a speaker voice $i$, which carries their biometric voiceprint, and the spoken text $t_i$. Standard anonymization treats the two as independent and perturbs only the first~\citep{EURECOM+7703}.
The mechanism acts on the speaker representation, and LDP requires that for any two
speakers, its output be $\eps$-indistinguishable, while the text passes through untouched. The flows are not, however, fully independent since both the voice and the text (vocabulary, phrasing, etc.) are shaped by the identity~\citep{aggazzotti2024can, 11461508}. 

In our experiments in Section \ref{sec:results}, we observe that holding the text fixed across speakers requires more injected noise to reach a given level of empirical privacy than letting it vary. This is consistent with prior findings that x-vector-style speaker
embeddings encode part of the spoken content \citep{raj2019probing} and
with the same-text versus different-text evaluation of
\citet{11464033}. When a speaker's text varies across utterances, this leakage adds within-speaker variation of the embedding. The inflated within-speaker spread degrades verification, so the text itself supplies part of the obfuscation. Fixing the text across all utterances removes this source, leaving the mechanism to provide all protection on its own. We therefore treat the same-text setting as the worse setting between the two, because it isolates the protection due to embedding perturbation alone, yielding a conservative estimate of what the mechanism provides.

\subsection{Laplace Mechanism}
We use a radial mechanism~\citep{hardt2010geometry, wu2017bolt}, calibrated with $\ell_2$-sensitivity. A coordinate-wise mechanism~\citep{dwork2016calibrating}, which adds independent Laplace noise per coordinate and is calibrated with $\ell_1$-sensitivity, behaves similarly throughout our experiments and is presented in Appendix~\ref{appx:coord}.
The mechanism operates directly within the speaker embedding space, injecting noise onto the high-dimensional embedding vectors. Because synthesis conditions only on the privatized embedding and the transcript, the released audio is a post-processing of the privatized embedding and inherits its $\eps$ guarantee. 

\begin{definition}[Radial Laplace mechanism~\citep{hardt2010geometry,wu2017bolt}]
\label{def:radial}
Let $\mathcal{E}:\mathcal{X} \rightarrow \mathbb{R}^{d}$ be a speaker encoder and $\mathcal{X}$ the set of utterances, with
$\ell_2$-sensitivity
$\Delta_2 = \max_{\mathbf{x},\mathbf{x}'\in\mathcal{X}}\lVert
\mathcal{E}(\mathbf{x})-\mathcal{E}(\mathbf{x}')\rVert_2$ and $\beta = \eps / \Delta_2$. The mechanism
\[
\mathcal{A}(\cdot )=\mathcal{E}(\cdot)+Z,\qquad
p_Z(z)\;\propto\;\exp\!\bigl(-\beta\lVert z\rVert_2\bigr),
\]
satisfies $\eps$-LDP for any $\mathbf{x} \in \mathcal{X}$.
\end{definition}

\begin{remark}[Sampling from Radial Laplace \citep{wu2017bolt}]
The radial Laplace mechanism uses $\Delta_2$ and adds a single isotropic perturbation $\widetilde{\mathcal{E}} = \mathcal{E} + R\,U$, where the direction $U = g/\lVert g\rVert_2$ is set by $g \sim \mathcal{N}(0,I_d)$ and the radius is drawn independently as $R \sim \mathrm{Gamma}(d,\,\varepsilon/\Delta_2)$.
\end{remark}

\begin{remark}
We report the empirical signal-to-noise ratio
$\mathrm{SNR} = \mathbb{E}\|\mathcal{E}(x)\|_2^2 / \mathbb{E}\|Z\|_2^2$ of the perturbed
embeddings; closed forms for both mechanisms are given in Appendix~\ref{appx:snr}.
\end{remark}

\renewcommand{\algorithmicrequire}{\textbf{Input:}}
\renewcommand{\algorithmicensure}{\textbf{Output:}}

\subsection{Exponential Pool Selection}
\label{sec:mechanism}

We construct a pool of candidates $\mathcal{P}$ from public donor speakers. The pool by itself does not leak any personal information since it is generated from an independent public donor dataset. However, publishing the candidate from the pool who best approximates the real speaker would violate privacy. To overcome this, we instead employ private selection using the exponential mechanism~\citep{mcsherry2007mechanism}. Given a source embedding $u\in\mathbb{R}^d\setminus\{0\}$ and a pool $\mathcal{P}=\{c_1,\dots,c_k\}$, we score each candidate by its cosine similarity to the source, $s_j = \cos(c_j,u) = \langle \hat{c}_j, \hat{u}\rangle$ where $\hat{c}_j = c_j/\lVert c_j\rVert_2$ and $\hat{u} = u/\lVert u\rVert_2$, and release a candidate index $j^\star$ sampled with probability proportional to $\exp(\lambda\, s_j)$. The parameter $\lambda$ controls the privacy-utility tradeoff: as $\lambda\!\to\!0$ selection is uniform over the pool, while large $\lambda$ concentrates mass on the candidates most similar to $u$. The released embedding is the selected candidate $c_{j^\star}$ at its original norm; the scoring normalization is used only to compute the selection distribution and does not alter the returned vector. 

Importantly, the released embedding is the selected donor, not an estimate of the source identity. Even the nearest donor within the finite pool need not be globally close to the source. Source information can instead leak through the source-dependent donor-selection probabilities, which the exponential mechanism bounds under DP.

\begin{algorithm}[H]
\caption{Exponential Pool Selection}
\label{alg:pool}
\begin{algorithmic}[1]
\Require source embedding $u\in\mathbb{R}^d\setminus\{0\}$; pool $\mathcal{P}=\{c_1,\dots,c_k\}$; budget $\varepsilon$; sensitivity $\Delta=2$
\Ensure anonymized embedding $c_{j^\star}$, at its original norm
\State $\lambda \gets \varepsilon/(2\Delta)$
\State $\hat{u} \gets u/\lVert u\rVert_2$
\For{$j \gets 1$ \textbf{to} $k$}
    \State $\hat{c}_j \gets c_j/\lVert c_j\rVert_2$
    \State $s_j \gets \langle \hat{c}_j,\hat{u}\rangle$
    \State $w_j \gets \exp(\lambda\, s_j)$
\EndFor
\State sample $j^\star$ with $\Pr[j^\star=j] = w_j\big/\sum_{i=1}^{k} w_i$
\State \Return $c_{j^\star}$
\end{algorithmic}
\end{algorithm}
\vspace{-3mm}

The privacy guarantee rests on the sensitivity of the score. Because each
candidate direction is a unit vector, a candidate's score can shift by at most
the distance between the normalized sources.
For any $u,u'$,
\[
  \big|s_j(u)-s_j(u')\big|
  = \big|\langle \hat{c}_j,\, \hat{u}-\hat{u}'\rangle\big|
  \le \lVert \hat{u}-\hat{u}'\rVert_2 \le 2,
\]
by Cauchy--Schwarz and $\lVert\hat c_j\rVert_2=1$, so the score sensitivity is
$\Delta=2$, the diameter of the unit sphere. Setting
$\lambda=\varepsilon/(2\Delta)=\varepsilon/4$ then calibrates the mechanism to
the privacy budget. The map from the selected index to its embedding is data-independent, so releasing $c_{j^\star}$ is
a post-processing of the private index and inherits its guarantee; the
downstream synthesizer, which depends on $u$ only through $c_{j^\star}$, does
likewise. In Appendix~\ref{appx:A}, we prove the following theorem.

\begin{theorem}{\cite{mcsherry2007mechanism}}
\label{thm:thm1}
Algorithm~\ref{alg:pool} is $\varepsilon$-LDP.
\end{theorem}

Theorem~\ref{thm:thm1} covers one run of Algorithm~\ref{alg:pool}, and we can run it at two different levels. At the \emph{speaker level}, we run it once per source speaker on the embedding of a fixed utterance from that speaker, and reuse the selected donor for all of that speaker's utterances. Reuse reveals nothing new, so each speaker costs $\varepsilon$ in total and always maps to the same donor. This requires a speaker identifier that is known in advance, such as a dataset label, which cannot be inferred from the audio. At the \emph{utterance level}, used when no such identifier exists (e.g., anonymizing unknown speakers in the wild), we select a fresh donor for every utterance. The guarantee then holds per utterance, and a speaker’s utterances may map to different donors.

\section{Evaluation}

\paragraph{Automatic Speaker Verification}
Since our goal is to render speech data safe for private use, we evaluate privacy preservation under an Automatic Speaker Verification (ASV) protocol \citep{EURECOM+7703}, adapting the two evaluation tasks of VoxGuard~\citep{11464033} (\texttt{User-Privacy-Strict} and \texttt{User-Privacy-Relaxed}). The task is to determine whether a probe utterance and an enrollment utterance (or set of $L$ utterances) were produced by the same speaker. The adversary first constructs a reference representation of the enrolled speaker from the enrollment set and then compares it against the probe utterance. Based on the resulting similarity score, it decides whether the probe belongs to the enrolled identity.

\begin{figure*}[t]
    \centering
    \begin{subfigure}[t]{0.49\linewidth}
        \centering
        \includegraphics[width=\linewidth]{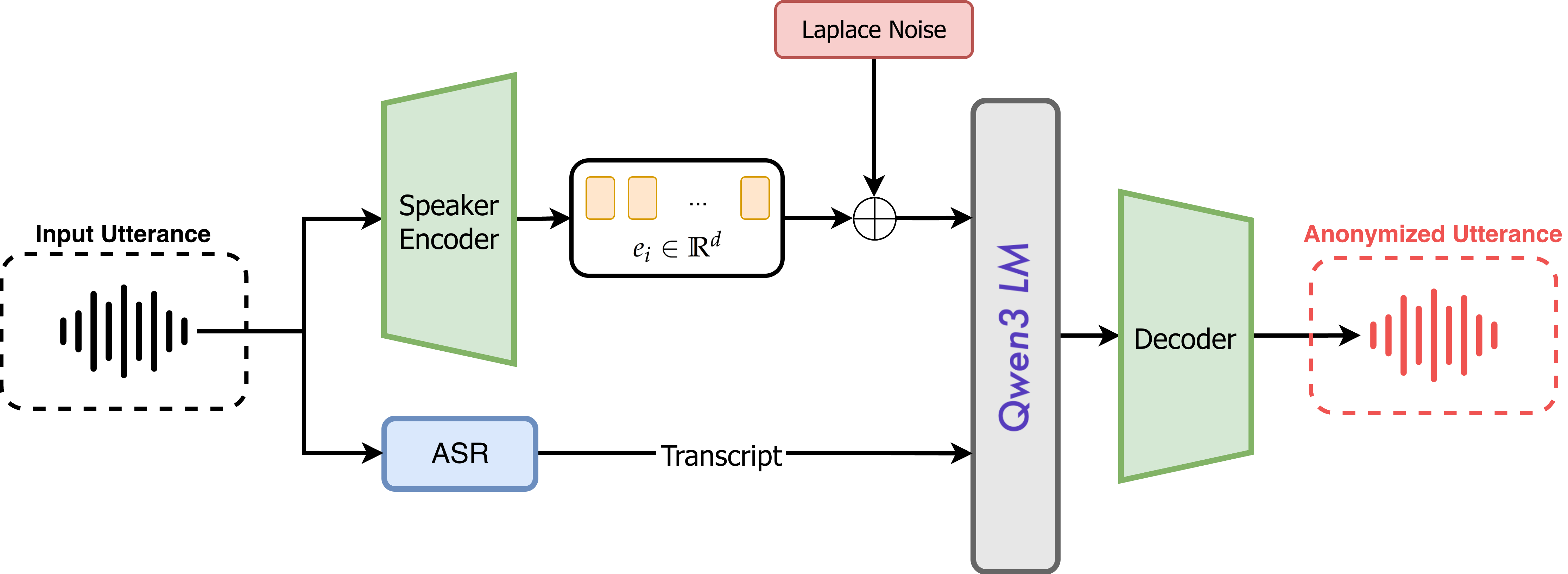}
        \caption{Additive Noise}
        \label{fig:anonymization-pipeline}
    \end{subfigure}
    \hfill
    \begin{subfigure}[t]{0.49\linewidth}
        \centering
        \includegraphics[width=\linewidth]{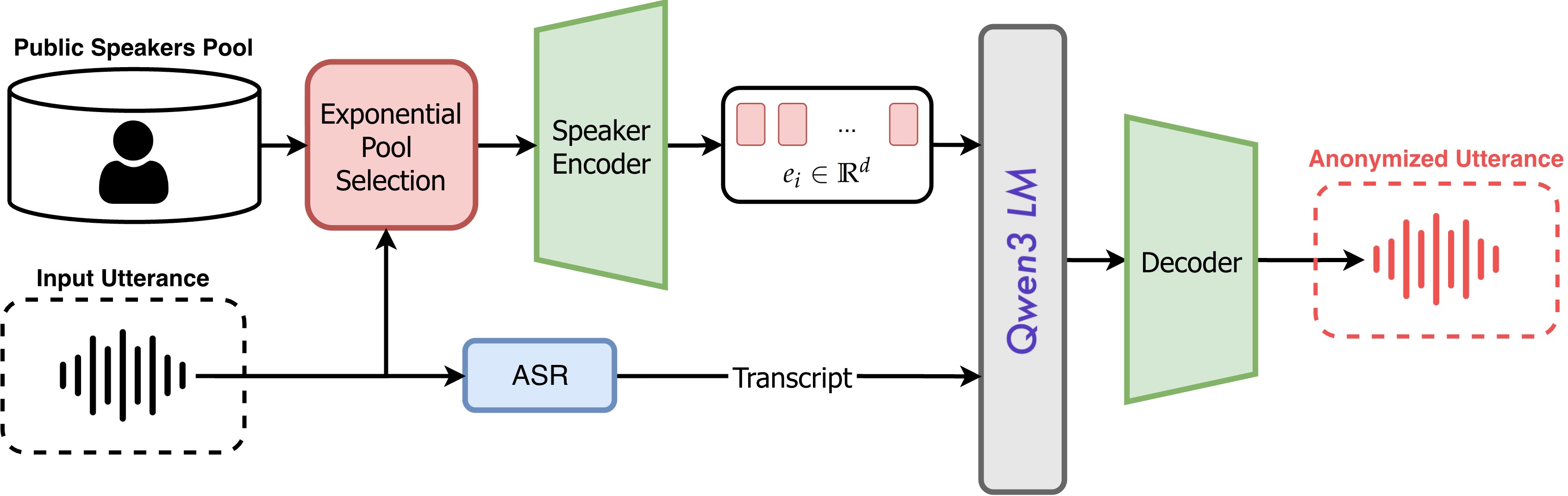}
        \caption{Pool Selection}
        \label{fig:anonymization-second}
    \end{subfigure}
    \caption{The two proposed anonymization mechanisms. Both extract a speaker embedding with ECAPA-TDNN and a transcript with Whisper, and resynthesize speech with Qwen3-TTS conditioned on a privatized embedding. (a)~\emph{Additive Noise} perturbs the source embedding with calibrated radial Laplace noise; (b)~\emph{Pool Selection} instead replaces it with a public donor speaker chosen from a fixed public pool via the exponential mechanism.}
    \label{fig:anonymization}
    \vspace{-5mm}
\end{figure*}


\paragraph{Anonymization}
We follow existing works that extract a speaker embedding, perturb it, and resynthesize speech from the perturbed representation~\citep{EURECOM+7703, meyer22b_interspeech,wang24ha_interspeech, 10096607, 10022601}. Specifically, given a reference utterance from speaker $i$, a speaker embedding $e_i \in \mathbb{R}^{2048}$ is produced by the ECAPA-TDNN \citep{desplanques20_interspeech} encoder jointly trained inside Qwen3-TTS \citep{hu2026qwen3}, and a transcript is obtained via an ASR front-end.  We operate the synthesizer in \texttt{x\_vector\_only\_mode}, discarding the in-context reference codec tokens and conditioning generation solely on $e_i$ and the transcript. This choice is essential, when in-context acoustic tokens are present they dominate speaker identity, making noise added to the embedding ineffective. Removing them makes $e_i$ the unique identity-carrying input, so that injecting noise $Z$ after the encoder satisfies the post-processing closure of DP. We privatize $e_i$ in one of two ways: by injecting noise $Z$ via the mechanism of Definition~\ref{def:radial}, or its coordinate-wise counterpart (Appendix~\ref{appx:coord}); or by replacing it with a donor embedding selected from the pool via Algorithm~\ref{alg:pool}. For speaker-level pool selection, the mechanism is run only once per speaker and the selected donor is cached in a lookup table indexed by an external, non-audio-derived identifier (e.g. a sequential ID). Subsequent utterances reuse the cached donor through deterministic lookup, which is post-processing and incurs no additional privacy cost. We assume the lookup table itself remains private. Both pipelines are shown in Figure~\ref{fig:anonymization}.

\paragraph{Threat Model}
We consider two adversaries that differ only in how they enroll the target speaker~\citep{williams2024anonymizing}. The \textbf{ignorant} adversary has no access to the anonymization system. It enrolls with original speech and compares it against anonymized probes. This is the weakest attacker, so privacy against it is necessary but not sufficient. The \textbf{lazy-informed} adversary knows the mechanism and its privacy budget $\varepsilon$, and anonymizes its enrollment utterances with the same mechanism at the same $\varepsilon$, so that enrollment and probe are anonymized under matched conditions. It draws its own randomness, does not observe the noise or donor drawn for the probe, and does not retrain its verifier on anonymized speech. Both adversaries use the same verifier, an ECAPA-TDNN model~\citep{desplanques20_interspeech} pretrained on original speech that extracts $192$-dimensional speaker embeddings, following VPC~\citep{EURECOM+7703} and VoxGuard~\citep{11464033}. A probe is scored by cosine similarity against the enrollment embedding; for $L>1$, the enrollment embeddings are averaged first. Both adversaries use only the acoustic channel and never inspect the transcript, as is standard in speaker-anonymization evaluation, where linguistic content enters as a utility measure (WER) rather than an attack surface~\citep{EURECOM+7703}.

\paragraph{Trials}
We build evaluation trials from the VCTK dataset~\citep{yamagishi2019vctk}, which comprises 109 speakers with approximately 400 utterances per speaker. Each trial consists of a probe utterance and an enrollment set containing $L$ utterances from a single enrolled speaker. A trial is \emph{positive} if the probe and enrollment belong to the same speaker and \emph{negative} otherwise. For each task and adversary, we construct 2,000 trials, balanced equally between positive and negative pairs, and report results for $L \in \{1,5\}$.
The two evaluation tasks differ in how the linguistic content is controlled. In \texttt{User-Privacy-Relaxed}, enrollment utterances are sampled randomly from the enrolled speaker and therefore contain diverse text. In \texttt{User-Privacy-Strict}, following Section~\ref{sec:privacy}, the probe and all enrollment utterances are constrained to have identical transcripts. This removes linguistic variability and isolates identity leakage to the acoustic channel. For positive trials, the enrollment therefore contains utterances with the same text as the probe and, in the worst case, may even contain the probe recording itself.

\paragraph{Pool Construction} We build the donor pool from  ParaSpeechCaps~\citep{diwan-etal-2025-scaling} that contains 594 speakers. For each donor speaker, we draw one utterance uniformly at random and embed it, yielding a pool of $k = 527$ candidate vectors (after filtering), one per speaker. No donor appears among the speakers being protected.

\paragraph{Sensitivity estimation} 

To measure sensitivity we impose a fixed, data-independent bound by clipping every speaker embedding to a ball of radius $R=18.5$, so that the $\ell_2$-sensitivity is the diameter of that ball, $\Delta_2 = 2R = 37$. However, speech embeddings are far from filling this ball. The observed diameter is only $\Delta_2 = 7.18$ on VCTK~\citep{yamagishi2019vctk} (the corpus we release) and $\Delta_2 = 10.74$ on the combined VCTK$+$ParaSpeechCaps~\citep{diwan-etal-2025-scaling} set. So the worst-case bound is $5.15\times$ the VCTK diameter and $3.44\times$ the combined one. This gap reflects that real speaker embeddings occupy a compact region of the manifold rather than spanning the full ball.  For the pool-of-speakers mechanism, the exponential mechanism uses cosine similarity as its scoring function, whose sensitivity is $\Delta  = 2$ since cosine values lie in $[-1,1]$.

\paragraph{Privacy Metrics}
We measure privacy with the Equal Error Rate (EER), following standard practice in speaker anonymization~\citep{EURECOM+7703}. Since EER can conceal worst-case leakage, we additionally report TPR@low-FPR, which measures worst-case verification with high adversary confidence~\citep{11464033}.

\paragraph{Utility}
We assess utility along four axes. \textbf{Content and quality}: Word Error Rate (WER) against a reference transcript measures linguistic preservation, and UTMOSv2~\citep{baba2024utmosv2} is a learned proxy for mean opinion score (MOS), predicting perceived speech naturalness on a 1--5 scale. Higher values indicate more natural and human-like speech, with 5 corresponding to near-human quality and 1 to severely degraded speech. \textbf{Consistency}: a useful anonymization mechanism should map each source speaker to a stable speaker from the donor pool. We quantify this via the gap between intra-speaker and inter-speaker cosine similarity of anonymized outputs, averaged across the dataset~\citep{miao2025benchmark}. \textbf{Downstream tasks}: we verify that anonymization preserves task-relevant information via intent classification and emotion recognition. 
For intent classification, we use the SLURP dataset~\citep{slurp}, evaluating whether anonymized speech preserves spoken language understanding via ASR transcription followed by intent, scenario, and action classification. For emotion recognition, we use the IEMOCAP dataset~\citep{busso2008iemocap}, classifying utterances into four emotions (angry, happy, sad, neutral) and reporting unweighted average recall (UAR)~\citep{feng22b_interspeech, EURECOM+7703}.

\subsection{Implementation Details}
We use the 1.7B-parameter variant of Qwen3-TTS and conduct all experiments on a single NVIDIA L40 GPU. Inference uses a batch size of 16, requiring approximately 6GB of GPU memory and 7 seconds per batch on average. Speaker embeddings for evaluation are extracted using the open-source ECAPA-TDNN model from SpeechBrain~\citep{ravanelli2021speechbrain}, pretrained on VoxCeleb~\citep{nagrani17_interspeech}. For automatic speech recognition, we employ the Whisper Base model~\citep{radford2023robust}.

\begin{figure}[h]
    \centering
    \includegraphics[width=0.98\linewidth]{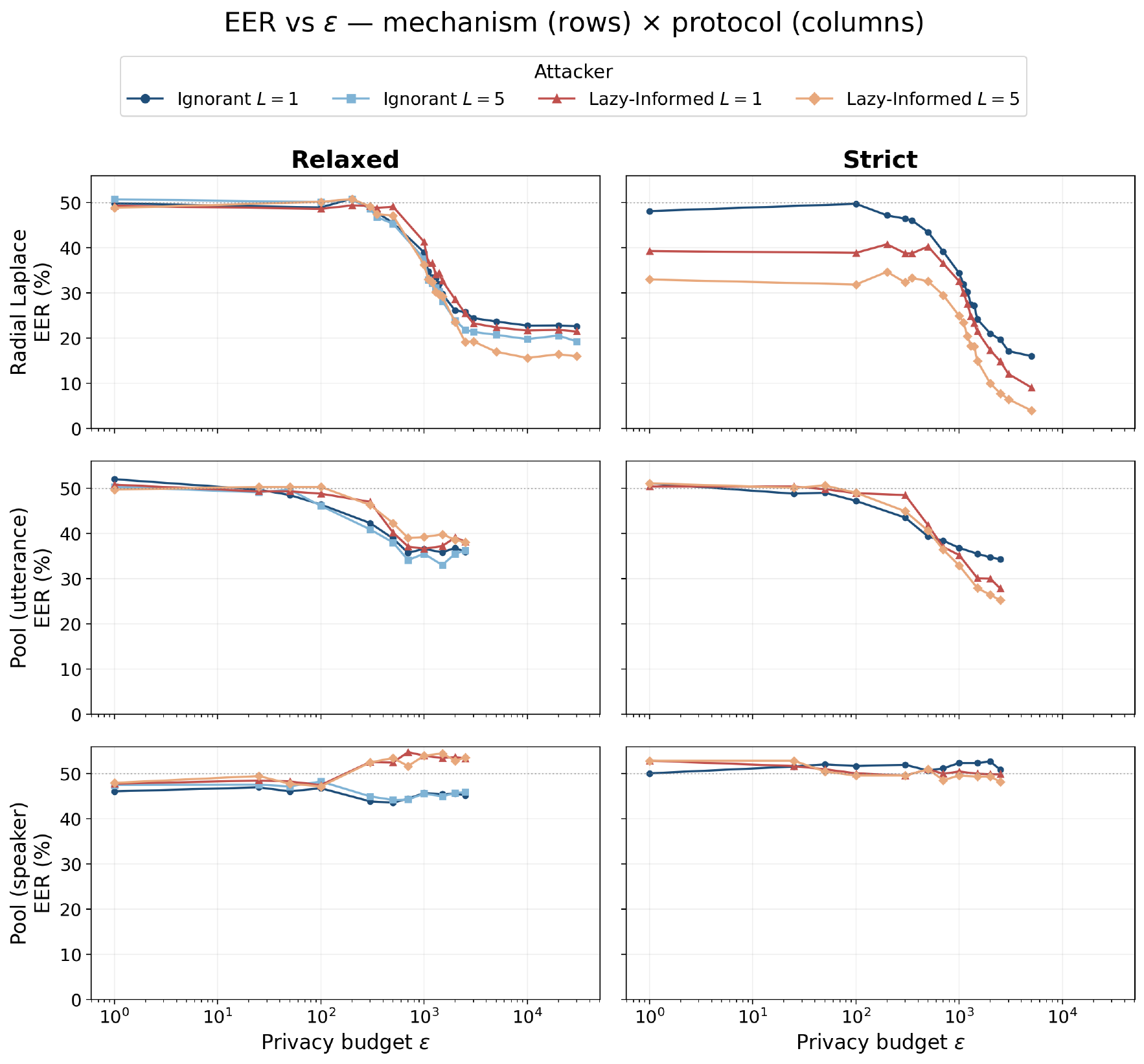}
    \caption{Empirical Privacy-$\eps$ curves for strict and relaxed cases under Radial Laplace and Pool Selection}
    \label{fig:eer}
    \vspace{-5mm}
\end{figure}

\section{Results}\label{sec:results}

\subsection{Operating Points for DP}

Figure~\ref{fig:eer} shows empirical privacy, measured by EER, as a function of the privacy budget $\varepsilon$.

\paragraph{Additive noise}
DP guarantees are strongest at small budgets ($\varepsilon \approx 1$), but the radial mechanism operates far from this regime. EER responds to $\varepsilon$ only above $\varepsilon \approx 2600$ and is flat below it. The embedding-space SNR explains the scale. An SNR of $1$ corresponds to $\varepsilon \approx 4700$, and the plateau begins once the SNR falls below about $0.3$, where the noise carries roughly three times the energy of the embedding. We hypothesize that the plateau reflects synthesizer failure rather than indistinguishability between speakers. Below the knee, the noise pushes the embedding outside the distribution on which Qwen3-TTS was trained, and the synthesizer often fails to emit an end-of-sequence token, decoding generic speech toward its maximum generation length. Under this account, \textit{the privacy observed in this regime comes from degraded synthesis, not from identity obfuscation}. The relevant operating point is therefore the knee of the curve: the largest budget at which EER is still at its plateau, and hence the least noise that attains it. The coordinate-wise mechanism behaves similarly throughout (Appendix~\ref{appx:coord}).

\paragraph{Pool selection}
Pool selection removes this failure mode since the synthesizer is always conditioned on a real donor embedding and never receives an out-of-distribution input. The mechanism can be applied at two granularities. At the \emph{utterance level}, a donor is drawn independently for every utterance, so no speaker identifier is needed. At the \emph{speaker level}, a donor is drawn once per speaker and reused for all of that speaker's utterances, which requires an external identifier. Up to $\varepsilon \approx 100$ the two behave alike: selection is close to uniform over the pool and EER stays at chance, at a budget more than an order of magnitude below the knee of the additive mechanisms. Beyond $\varepsilon \approx 100$ they diverge. At the utterance level, each draw concentrates on the donors nearest the source, so a speaker's utterances map to the same few donors and become linkable; EER falls to ${\approx}26\%$ under strict evaluation. At the speaker level, EER stays near chance over the entire range. Pool size matters only at the utterance level, since a larger pool spreads a speaker's utterances over more donors and raises EER, while speaker-level EER stays near chance at every pool size (Appendix~\ref{app:pool_size}). The formal guarantee is the same for every pool size.

 \begin{figure*}[!ht]
    \centering

    \begin{subfigure}{0.45\linewidth}
        \centering
        \includegraphics[width=\linewidth]{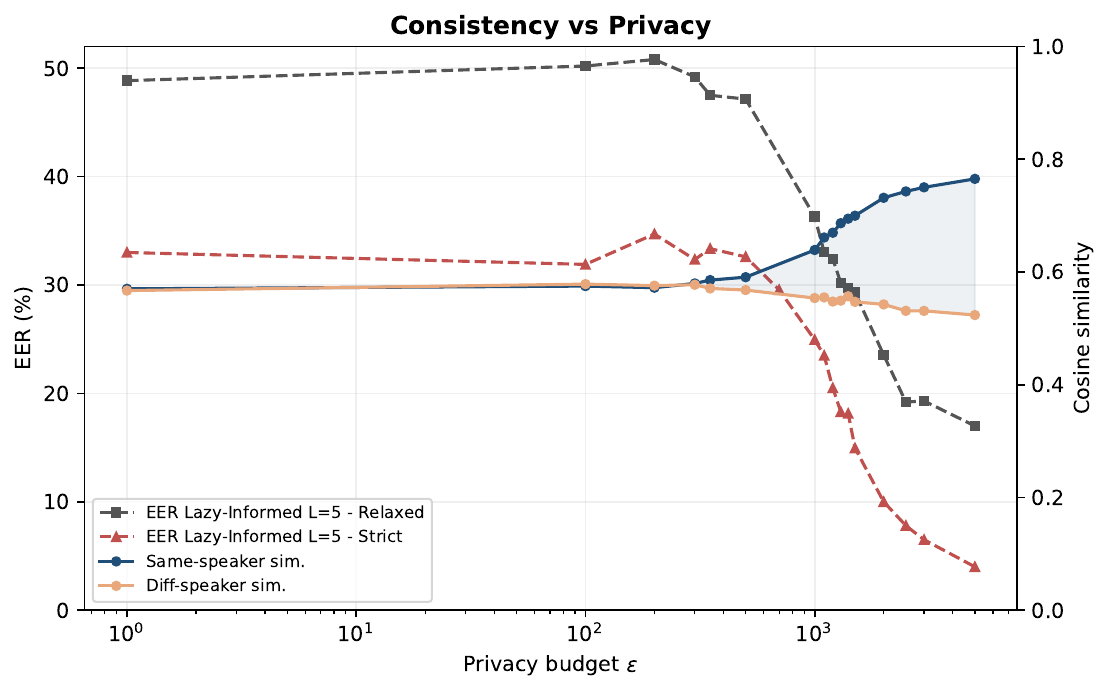}
        \caption{Radial Laplace}
        \label{fig:consistency_radial}
    \end{subfigure}
    \hspace{0.02\linewidth}
    \begin{subfigure}{0.45\linewidth}
        \centering
        \includegraphics[width=\linewidth]{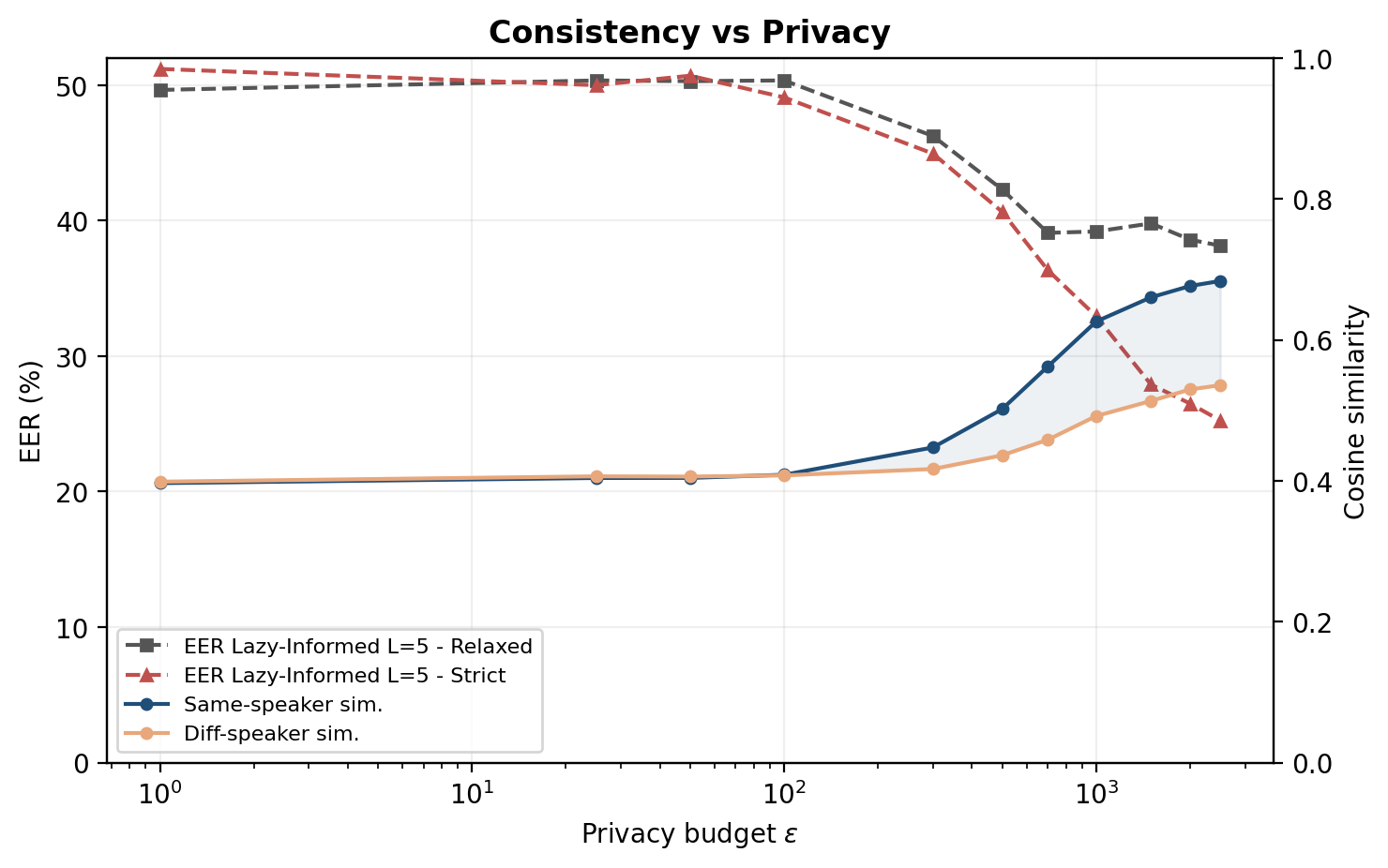}
        \caption{Utterance-level Pool Selection}
        \label{fig:consistency_pool_utt}
    \end{subfigure}
    \vspace{-2mm}
    \caption{Consistency of the anonymization mechanisms across privacy budgets.}
    \label{fig:consistency}
    \vspace{-3mm}
\end{figure*}

\subsection{Utility}

\paragraph{Content and quality}
Across all mechanisms, anonymization preserves speech quality, linguistic content, and spoken language understanding. WER remains below $0.08$ and UTMOS stays within $3.3$--$3.6$ across the full range of privacy budgets, while intent, scenario, and action accuracy on SLURP vary by less than $2.5$ percentage points from their no-noise baselines. Perturbing the speaker representation therefore leaves semantic content and perceptual quality largely unaffected.

\paragraph{Emotion}
Emotion recognition is the only downstream task with a meaningful privacy--utility trade-off. Under radial Laplace, stronger privacy systematically reduces UAR. We observe that accuracy on expressive emotions drops while neutral speech becomes easier to recognize, so the perturbation progressively neutralizes the speech signal. Under pool selection, UAR remains largely stable across privacy budgets, because the mechanism samples natural speaker embeddings instead of perturbing them. Detailed WER, UTMOS, SLURP, and SER results, including the per-emotion analysis, are in Appendix~\ref{appx:additional}.

\paragraph{Consistency}
Stronger privacy also costs consistency. An anonymization system should map each source speaker to a stable surrogate across utterances. Under radial Laplace (Figure~\ref{fig:consistency_radial}), the gap between intra- and inter-speaker similarity narrows as privacy increases, and once the EER plateau is reached, utterances from the same speaker are no more similar than utterances from different speakers. Utterance-level pool selection loses consistency in two ways (Figure~\ref{fig:consistency_pool_utt}). At large budgets, selection collapses to only 18 of the 527 donors, so many unrelated source speakers map to the same few donors. For $\varepsilon < 100$, selection is approximately uniform over the pool, so a speaker's utterances are assigned donors essentially at random. Speaker-level pool selection avoids both problems since it runs the exponential mechanism once per speaker and reuses the same donor for all of that speaker's utterances, which maintains a consistent speaker-to-speaker mapping.

\subsection{Does a More Informed Attacker Help?}

Figure~\ref{fig:eer} compares the attackers under both text protocols.

\paragraph{Additive noise}
In the \emph{relaxed} setting, knowing the mechanism helps little: the lazy-informed attacker stays within $1$--$2$ points of the ignorant one over most of the range where EER responds to $\varepsilon$, and the gap peaks at ${\approx}5$ points near $\varepsilon \approx 1000$. In the \emph{strict} setting, enrollment and probe share a transcript, so the residual similarity is attributable to the acoustic channel alone. EER is lower at every budget, and the lazy-informed attacker gains a further $5$--$10$ points across the transition region. \textit{Matching the transcript therefore exposes leakage that the relaxed setting largely hides.} Enrollment size compounds the effect. Relative to $L=1$, $L=5$ lowers EER by $2$--$5$ points (relaxed) and $6$--$8$ points (strict); near $\varepsilon = 500$, the strict lazy-informed attacker falls from ${\approx}40\%$ to ${\approx}33\%$. A larger enrollment set gives the adversary several independently privatized releases of the same identity, which is \textit{composition}, and each additional release weakens the effective guarantee.

\paragraph{Pool selection}
Attacker knowledge again yields little in the relaxed setting, where the configurations differ by at most $3$ points. The protocols differ instead in the \emph{shape} of the transition. At the utterance level, the strict curves hold at chance until $\varepsilon \approx 500$ and then fall steeply to ${\approx}26\%$, whereas the relaxed curves decline gradually and reach only ${\approx}36\%$ over the same range. We again attribute the difference to the fixed transcript, which keeps content mismatch from diluting the identity signal. At the speaker level, one map governs all of a speaker's utterances, so the mechanism is invoked once regardless of $L$ and there is no composition to observe. Accordingly, the $L=1$ and $L=5$ curves coincide under both protocols.

\subsection{Beyond Average-Case Privacy}\label{sec:tpr}

The low-FPR analysis reveals leakage that EER alone does not show.

\paragraph{Additive noise}
Even in the \emph{relaxed} setting, the advantage of the lazy-informed attacker becomes visible once $L=5$, reaching ${\approx}8\%$ $\mathrm{TPR}@1\%\mathrm{FPR}$ at $\varepsilon \approx 5000$. The \emph{strict} setting exposes substantially stronger worst-case leakage. Near the knee of the curve ($\varepsilon \approx 2500$), the lazy-informed attacker still achieves $20$--$30\%$ $\mathrm{TPR}@1\%\mathrm{FPR}$, so a non-trivial fraction of speakers remains open to highly confident verification. The gap between the two settings is most pronounced at smaller budgets: at $\varepsilon \approx 1500$, the strict attacker attains ${\approx}9\%$ $\mathrm{TPR}@1\%\mathrm{FPR}$, whereas the relaxed attacker has already dropped to ${\approx}1.6\%$.

\paragraph{Pool selection}
Both granularities remain at chance ($\mathrm{TPR} \approx \mathrm{FPR}$) across the entire budget range and under both protocols, and tightening to $0.5\%$ FPR does not change this. The contrast with additive noise is clear. There, EER understates the risk: near the knee, average-case performance approaches chance while a subpopulation remains confidently re-identifiable. For pool selection, no operating point yields high-confidence re-identification, so the attacker's advantage is spread across the population rather than concentrated in a vulnerable minority. The complete low-FPR results are in Appendix~\ref{appx:privacy}.

\section{Discussion}

We presented three DP mechanisms that operate on the speaker embedding and evaluated them under a protocol that controls for transcript content. Our findings concern identity carried by the acoustic channel, and we identify two directions beyond it. First, protecting attributes such as accent or age lacks both a formal definition and a means of evaluation. Reliable verifiers do not exist for every attribute, so how effectively a mechanism obfuscates a specific trait, and with what guarantee, remains open. Second, our mechanisms do not account for the fact that speakers are also identifiable through their linguistic traits~\citep{tsaprazlis2026sparc, aggazzotti2024can}. Modeling those traits, understanding how content transformations affect them, and determining the optimal anonymization against content-aware adversaries are necessary for any complete speech privacy mechanism.

\section{Conclusion}

We presented speaker anonymization mechanisms with provable DP guarantees, obtained either by injecting calibrated noise into the speaker representation or by releasing a donor speaker drawn from a fixed pool under the exponential mechanism. We characterized practical operating points for differential privacy in speech and identified a worst-case evaluation protocol that isolates leakage of acoustic identity. Pool selection achieves provable DP without the utility cost that excessive noise imposes on the synthesizer, which exposes the limits of simple noise injection in the embedding space. Applied at the speaker level, it also maintains consistent pseudo-identities across a speaker's utterances.

\section*{Limitations}

First, our guarantee covers the speaker embedding alone; the transcript is released unmodified. An adversary that also reads the transcript is outside the threat model, and since a speaker's vocabulary and phrasing are themselves identifying, our results do not bound the leakage available to one. Second, our empirical evaluation does not include a semi-informed attacker whose verifier is retrained on anonymized speech. Such an attacker may exploit residual artifacts of the anonymization pipeline and achieve higher linkage rates. Our empirical results are therefore a lower bound on attacker success; the formal DP guarantee does not depend on the evaluated attacker. Third, pool selection depends on the donor set. The selected donor can be no closer to the source than the pool's nearest member, so a finite pool imposes a floor on voice mismatch that no increase in $\varepsilon$ removes. We vary pool size only under the lazy-informed, different-text protocol (Appendix~\ref{app:pool_size}), measure source-surrogate similarity but not privacy on the interpolated $100$K pool (Appendix~\ref{app:similarity}), and do not vary the pool's demographic composition. Characterizing these dependencies is left to future work.

\section*{Ethical Considerations}

This work aims to make speech data safer to share by attaching formal privacy guarantees to speaker identity. We note two considerations.

\paragraph{Data}
All experiments use publicly released corpora under their original licenses: VCTK~\citep{yamagishi2019vctk}, ParaSpeechCaps~\citep{diwan-etal-2025-scaling}, SLURP~\citep{slurp}, and IEMOCAP~\citep{busso2008iemocap}. Donor speakers are drawn from a public corpus and are disjoint from the speakers our mechanism protects. They receive no privacy guarantee from this work, and the released surrogate voices are their real voices, not synthetic constructions. Where the source corpus's license does not cover such reuse of donor voices, deployments should populate the pool with synthetic voices, which the mechanism supports without modification.

\paragraph{Dual use}
Anonymization tools transfer directly to voice impersonation: the machinery that renders a speaker in a surrogate voice also renders an arbitrary transcript in that voice. This risk is inherent to any TTS system capable of speaker conditioning and is not introduced by our contribution, but our pipeline lowers the effort required to apply it at scale. Anonymized speech may also be used to circumvent voice-based authentication.

\section*{Acknowledgements}
This work was supported by the USC Amazon Center for Secure
and Trusted Machine Learning and the Office of the Director of National Intelligence (ODNI), Intelligence Advanced Research Projects Activity (IARPA), via the HIATUS Program under contract \#2022\-22072200006. The views and conclusions contained herein are those of the authors and should not be interpreted as necessarily representing the official policies, either expressed or implied, of ODNI, IARPA, or the U.S. Government. The U.S. Government is authorized to reproduce and distribute reprints for governmental purposes notwithstanding any copyright annotation therein.

\bibliography{custom}

\clearpage
\onecolumn 
\appendix

\section*{Appendix} \label{appx}

\section{Proof of Theorem \ref{thm:thm1} } \label{appx:A}

\begin{proof}
Let $Z_u=\sum_{c'\in\mathcal{P}}\exp(\lambda\langle c',u\rangle)$. Then
\[
  \frac{\Pr[c\mid u]}{\Pr[c\mid u']}
  =\exp\!\big(\lambda\langle c,\,u-u'\rangle\big)
   \cdot
   \frac{Z_{u'}}{Z_{u}}
\]
 All vectors  are unit-norm and by the Cauchy--Schwarz inequality we have: $\exp(\lambda\langle c,u-u'\rangle)\le \exp(\lambda \cdot \lVert c\rVert_2 \cdot \lVert u-u'\rVert_2) \leq  \exp(2 \lambda) $. Furthermore, for every element $c' \in \mathcal{P}$,
$\exp(\lambda\langle c',u'\rangle)=\exp(\lambda\langle c',u\rangle)\exp(\lambda\langle c',u'-u\rangle)
\le \exp(\lambda\langle c',u\rangle)\,\exp(\lambda\lVert u-u'\rVert_2) \leq \exp(2\lambda) \cdot \exp(\lambda\langle c',u\rangle)$. Summing:
\[
  Z_{u'}=\sum_{c'}\exp(\lambda\langle c',u'\rangle)
  \le \exp(2\lambda)\sum_{c'}\exp(\lambda\langle c',u\rangle)
  = \exp(2\lambda)\,Z_u,
\]
so $Z_{u'}/Z_u\le \exp(2\lambda)$. Therefore, $\frac{\Pr[c\mid u]}{\Pr[c\mid u']} \le \exp(4\lambda)
=\exp(\varepsilon)$ for $\lambda=\varepsilon/4$.
\end{proof}

\section{Signal-to-Noise Ratio}\label{appx:snr} 

We also evaluate our noise-injecting Laplace mechanisms by empirically calculating the Signal-to-Noise Ratio (SNR). By clipping, we know $\Delta_1,\Delta_2$ and we compute $\mathbb{E}\,\, \|\mathcal{E}(x)\|_2^2$ empirically and the second moment of the noise norm mathematically as follows:

For the coordinate and radial Laplace mechanisms, we have:

\[
\mathbb{E}\,\lVert Z_{\mathrm{coord}}\rVert_2^2
= 2d\!\left(\frac{\Delta_1}{\eps}\right)^{2},
\]
\[
\mathbb{E}\,\lVert Z_{\mathrm{rad}}\rVert_2^2
= d(d+1)\!\left(\frac{\Delta_2}{\eps}\right)^{2}
\]

The final SNR is $\mathbb{E}\,\|\mathcal{E}(x)\|_2^2 \,/\, \mathbb{E}\,\|Z\|_2^2$.

\section{Coordinate-wise Laplace}\label{appx:coord}

This section states the coordinate-wise Laplace mechanism in $d$ dimensions \citep{dwork2016calibrating}. Unlike
the radial mechanism of Definition~\ref{def:radial}, which adds a single isotropic perturbation calibrated to $\ell_2$-sensitivity, it perturbs each
coordinate independently and is calibrated to $\ell_1$-sensitivity. Empirical
results are reported in Appendix~\ref{appx:additional}.
\begin{definition}[Coordinate-wise Laplace mechanism]
Let $\mathcal{E}:\mathcal{X} \rightarrow \mathbb{R}^{d}$ be a speaker encoder and $\mathcal{X}$ the set of utterances, with 
$\ell_1$-sensitivity
$\Delta_1 = \max_{\mathbf{x},\mathbf{x}'\in\mathcal{X}}\lVert
\mathcal{E}(\mathbf{x})-\mathcal{E}(\mathbf{x}')\rVert_1$. Let $Z=(z_1,\dots,z_d)$ with
$z_i \sim \mathrm{Lap}(b)$ i.i.d.\ and $b = \Delta_1 / \eps$. Then
$\mathcal{A}(\cdot ) = \mathcal{E}(\cdot ) + Z$ satisfies $\eps$-LDP for any
$\mathbf{x}\in\mathcal{X}$.
\end{definition}

\section{Additional Results}\label{appx:additional}
\subsection{Privacy}\label{appx:privacy}

\begin{figure}[!ht]
    \centering
    \includegraphics[width=0.85\linewidth]{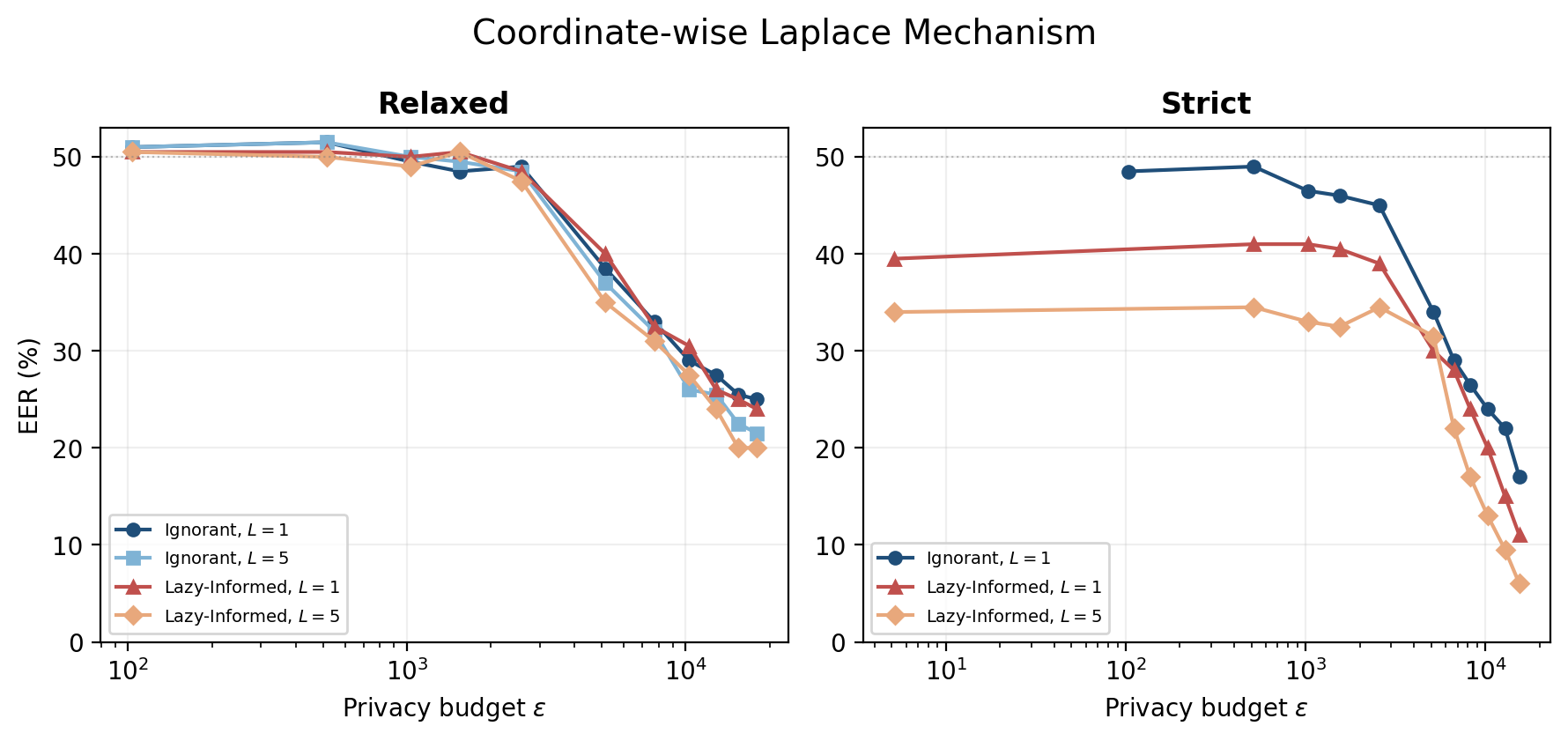}
    \caption{Empirical Privacy-$\eps$ curves for strict and relaxed cases under coordinate-wise Laplace}
    \label{fig:laplaceEER}
\end{figure}

\begin{figure}[!ht]
    \centering
    \includegraphics[width=0.8\linewidth]{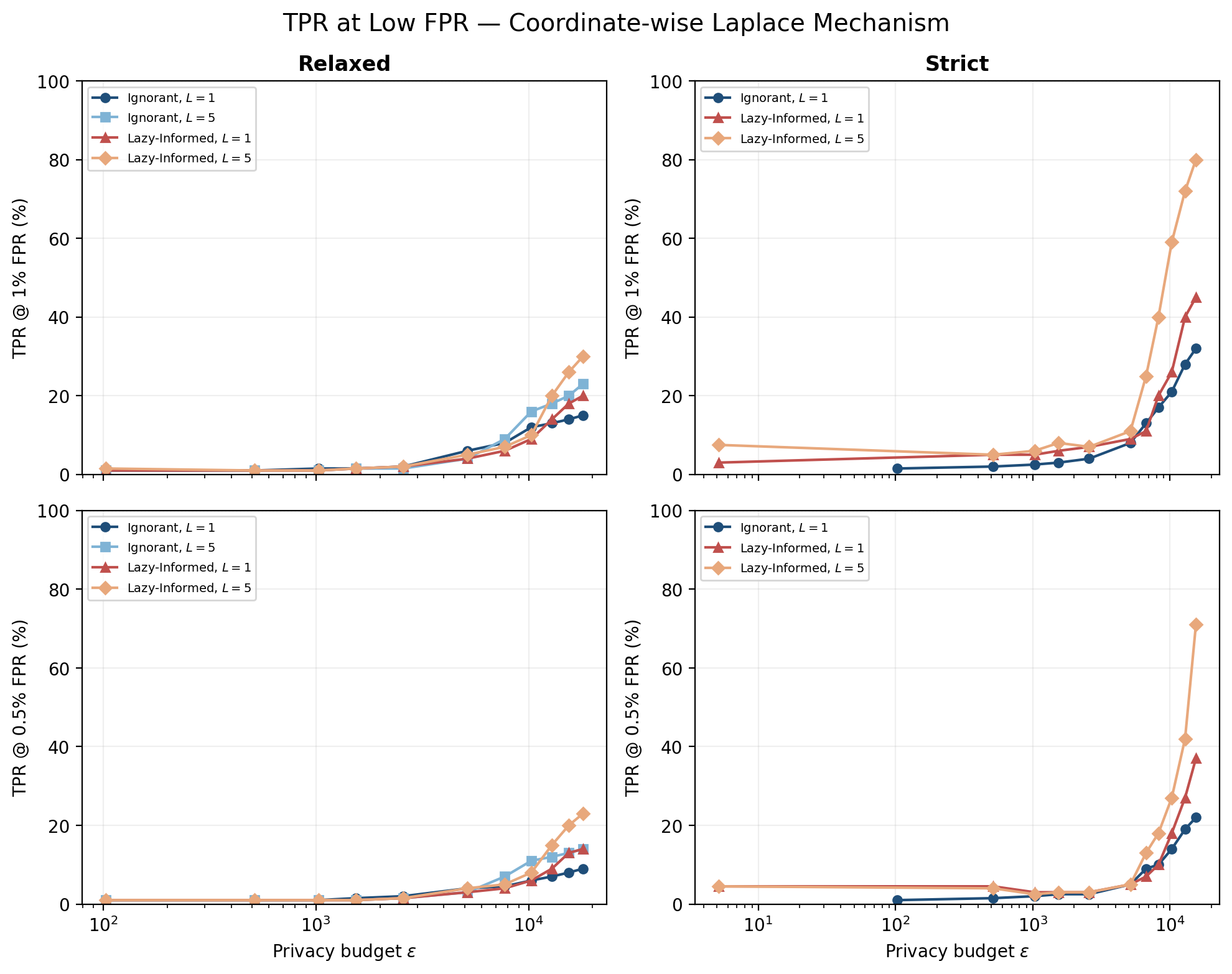}
    \caption{Low-FPR privacy evaluation for the coordinate-wise Laplace mechanism. Results are reported as $\mathrm{TPR}@\mathrm{FPR}$ for $\mathrm{FPR}\in\{1\%,0.5\%\}$ under strict and relaxed settings}
    \label{fig:laplaceLOWFPR}
\end{figure}

\begin{figure}[!ht]
    \centering
    \includegraphics[width=0.78\linewidth]{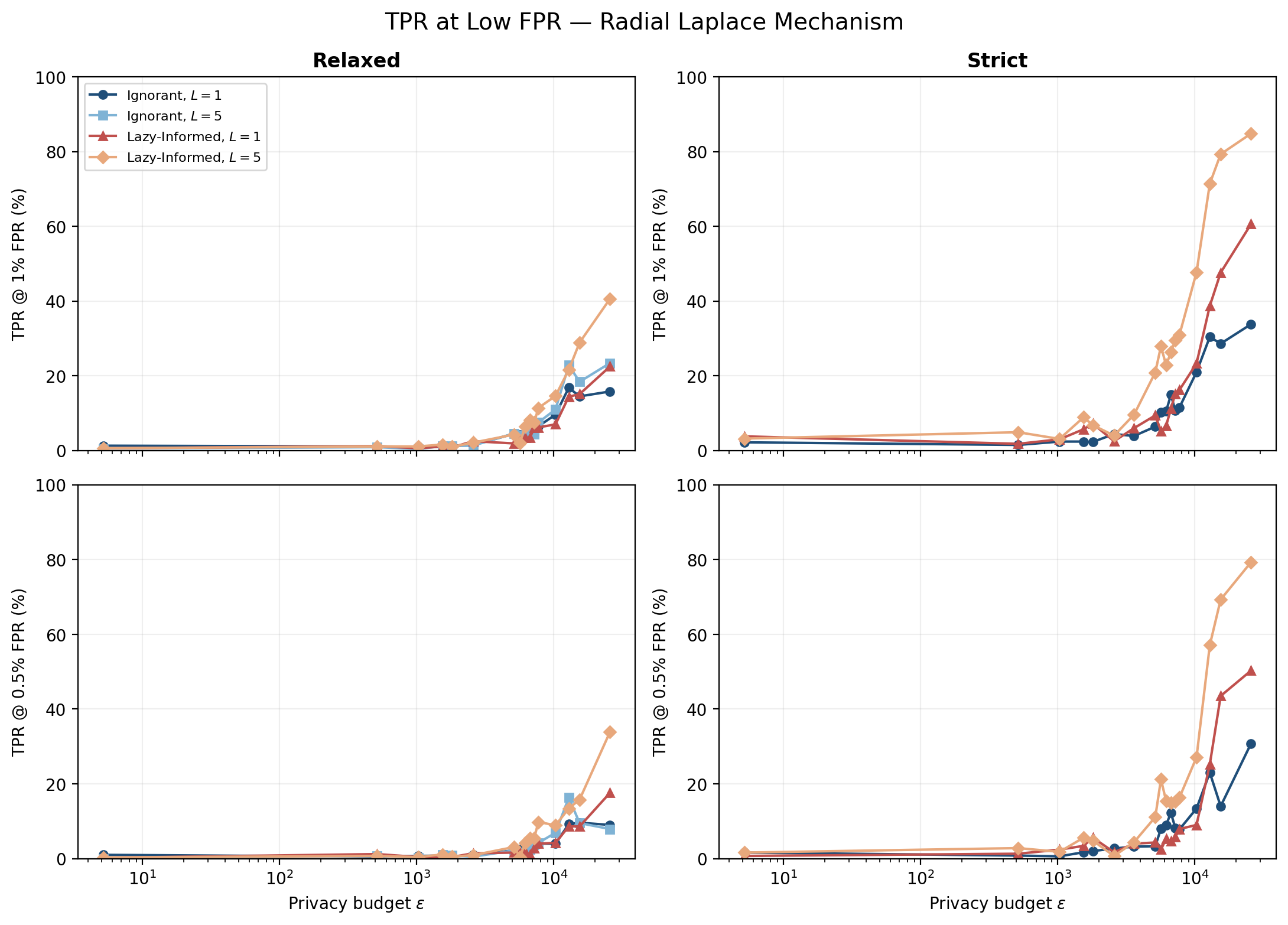}
    \caption{Low-FPR privacy evaluation for the radial Laplace mechanism. Results are reported as $\mathrm{TPR}@\mathrm{FPR}$ for $\mathrm{FPR}\in\{1\%,0.5\%\}$ under strict and relaxed settings}
    \label{fig:tpr}
\end{figure}

Figure~\ref{fig:laplaceEER} reports empirical privacy for the coordinate-wise mechanism. The qualitative picture matches the radial case in every respect, which is why we defer it here. Under \emph{relaxed} evaluation, all four attacker configurations sit at chance-level EER ($\approx 50\%$) up to $\varepsilon \approx 1500$ and then decline together, reaching $20$--$25\%$ at the largest budgets evaluated. Neither mechanism knowledge nor enrollment size separates the curves by more than $\approx 5$ percentage points anywhere in this range, and the four configurations are indistinguishable throughout the plateau. The \emph{strict} setting separates the attackers, and does so already at the plateau. The ignorant attacker saturates near $48\%$, the lazy-informed attacker at $\approx 40\%$ for $L=1$ and $\approx 34\%$ for $L=5$. Mechanism knowledge is worth $\approx 8$--$10$ points across the plateau and the transition region, and increasing enrollment from $L=1$ to $L=5$ a further $5$--$6$ points; both magnitudes agree with the radial mechanism. At the largest budgets, the ordering is preserved and the spread widens, with EER falling to $\approx 17\%$, $11\%$, and $6\%$ respectively. Figures~\ref{fig:laplaceLOWFPR} and~\ref{fig:tpr} report $\mathrm{TPR}@\mathrm{FPR}$ for $\mathrm{FPR} \in \{1\%, 0.5\%\}$ under both mechanisms. The two evaluation settings diverge more sharply here than under EER. At the largest budget evaluated, the lazy-informed attacker with $L=5$ attains $\approx 30\%$ $\mathrm{TPR}@1\%\mathrm{FPR}$ under relaxed evaluation but $\approx 80\%$ under strict evaluation for the coordinate-wise mechanism, and $\approx 49\%$ versus $\approx 85\%$ for the radial mechanism. This is the expected signature of a heavy-tailed per-speaker vulnerability distribution: a small subpopulation is verified with high confidence while the bulk of the population is not, and averaging over enrollment utterances removes precisely those cases from view.

\begin{figure}[!ht]
    \centering
    \includegraphics[width=0.72\linewidth]{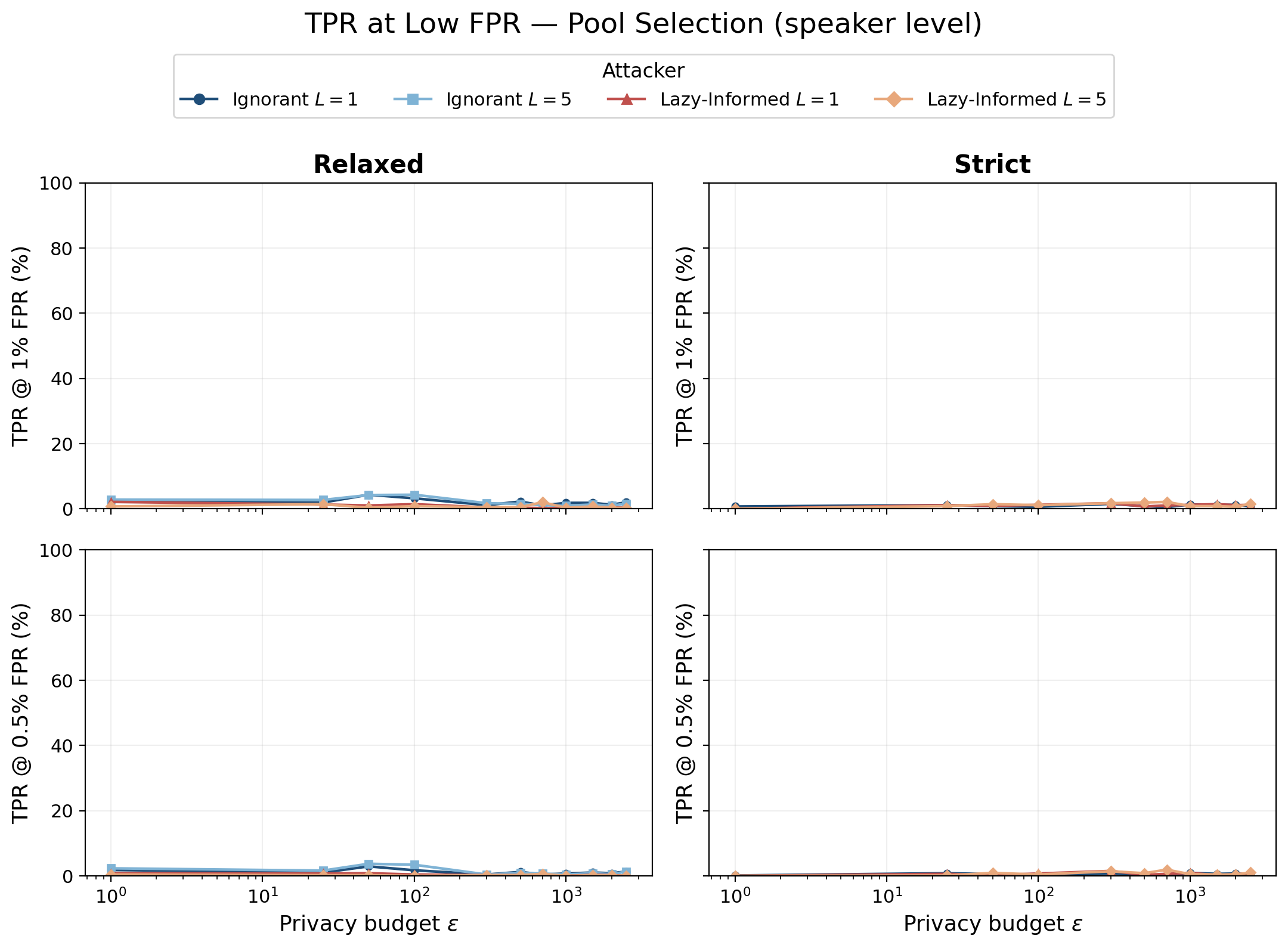}
    \caption{Low-FPR privacy evaluation for the exponential pool selection mechanism in the speaker granularity. Results are reported as $\mathrm{TPR}@\mathrm{FPR}$ for $\mathrm{FPR}\in\{1\%,0.5\%\}$ under strict and relaxed settings}
    \label{fig:poolLOWFPR}
\end{figure}

\begin{figure}[!ht]
    \centering
    \includegraphics[width=0.75\linewidth]{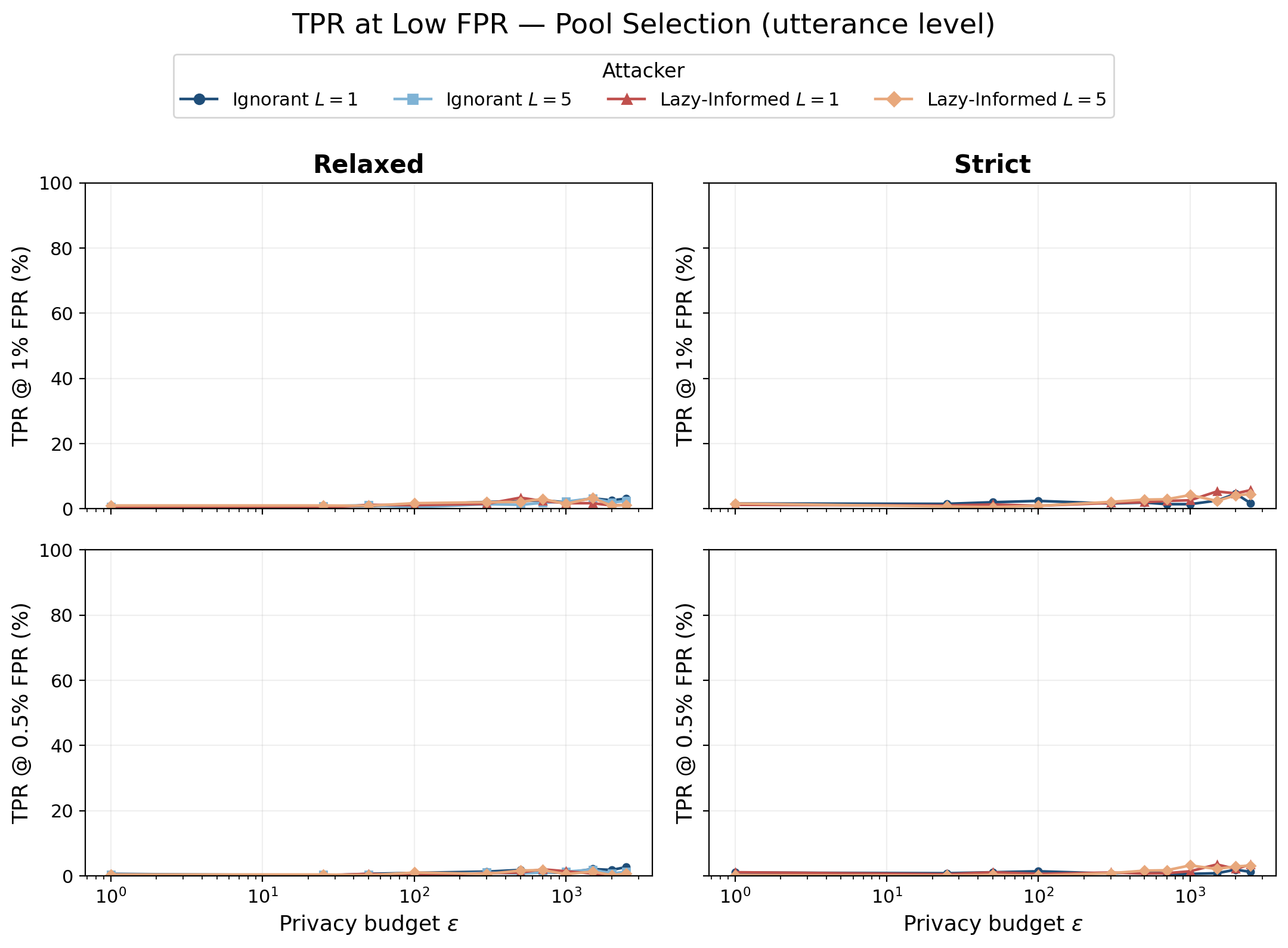}
    \caption{Low-FPR privacy evaluation for the exponential pool mechanism in the utterance granularity. Results are reported as $\mathrm{TPR}@\mathrm{FPR}$ for $\mathrm{FPR}\in\{1\%,0.5\%\}$ under strict and relaxed settings}
    \label{fig:poll_utt_tpr}
\end{figure}

\subsection{Pool Size Ablation}
\label{app:pool_size}

The privacy guarantee of Algorithm~1 holds for any fixed, source-independent public donor pool, but pool size can affect empirical privacy. We therefore repeat the lazy-informed, different-text evaluation with $k\in\{50,250\}$ donors, for both speaker- and utterance-level selection and enrollment sizes $L\in\{1,5\}$. Table~\ref{tab:pool_size} reports the complete EER results.

\begin{table*}[t]
\centering
\small
\begin{tabular}{r|cc|cc|cc|cc}
\toprule
& \multicolumn{4}{c|}{\textbf{Speaker-level}}
& \multicolumn{4}{c}{\textbf{Utterance-level}} \\
& \multicolumn{2}{c|}{$k=50$}
& \multicolumn{2}{c|}{$k=250$}
& \multicolumn{2}{c|}{$k=50$}
& \multicolumn{2}{c}{$k=250$} \\
$\varepsilon$
& $L=1$ & $L=5$
& $L=1$ & $L=5$
& $L=1$ & $L=5$
& $L=1$ & $L=5$ \\
\midrule
1    & 46.0 & 46.6 & 46.2 & 46.8 & 51.2 & 48.2 & 50.1 & 51.6 \\
25   & 49.1 & 48.4 & 46.8 & 48.1 & 52.1 & 50.3 & 48.8 & 53.0 \\
50   & 47.7 & 47.2 & 46.0 & 44.4 & 50.2 & 50.9 & 49.8 & 52.1 \\
100  & 48.6 & 49.1 & 53.4 & 54.0 & 51.6 & 52.5 & 48.5 & 51.8 \\
300  & 48.2 & 48.9 & 52.6 & 53.6 & 47.4 & 45.9 & 45.4 & 48.0 \\
500  & 47.5 & 46.9 & 53.7 & 53.1 & 42.1 & 40.5 & 40.4 & 44.3 \\
700  & 48.5 & 49.0 & 53.6 & 52.9 & 37.2 & 38.6 & 38.2 & 41.3 \\
1000 & 47.8 & 47.3 & 54.1 & 54.7 & 33.9 & 34.8 & 38.7 & 40.0 \\
1500 & 48.7 & 48.2 & 53.5 & 54.1 & 33.8 & 33.2 & 38.9 & 41.5 \\
2000 & 48.4 & 48.0 & 52.5 & 54.2 & 32.0 & 31.2 & 39.9 & 41.1 \\
2500 & 49.9 & 48.2 & 53.9 & 54.5 & 30.2 & 31.3 & 39.6 & 41.9 \\
\bottomrule
\end{tabular}
\caption{Pool-size ablation under the lazy-informed, different-text protocol. Values are EER (\%); higher indicates stronger empirical privacy. We compare pool sizes $k\in\{50,250\}$, speaker- and utterance-level selection, and enrollment sizes $L\in\{1,5\}$.}
\label{tab:pool_size}
\end{table*}

\paragraph{Speaker-level selection}
Speaker-level selection remains approximately at chance across both pool sizes, privacy budgets, and enrollment sizes. Algorithm~\ref{alg:pool} is invoked only once per source speaker and the selected donor is reused for all subsequent utterances. Consequently, increasing $L$ does not expose additional independently privatized realizations of the source speaker. The similar behavior for $k=50$ and $k=250$ further shows that the strong empirical privacy of speaker-level selection is not specific to the original pool size.

\paragraph{Utterance-level selection}
Pool size has a substantially larger effect at utterance-level granularity. At small $\varepsilon$, selection is approximately uniform and both pool sizes remain near chance. As $\varepsilon$ increases, donor selection becomes increasingly source-dependent and EER decreases. This decrease is weaker for the larger pool, particularly with $L=5$. For example, at $\varepsilon=2500$, EER increases from $31.3\%$ with $k=50$ to $41.9\%$ with $k=250$. This behavior is consistent with repeated independent donor selection. A larger pool provides more possible donor identities, reducing the probability of repeated or similar donor assignments across independently anonymized utterances from the same source. Multiple enrollment utterances therefore provide the attacker with a less consistent acoustic signal for speaker linking.

\paragraph{Privacy guarantee}
These differences concern empirical speaker-linking performance, not the formal privacy guarantee. Algorithm~1 remains $\varepsilon$-LDP for any fixed public pool that is independent of the private source: changing $k$ changes the output space and the empirical behavior, but not the sensitivity or the privacy calibration of the exponential mechanism. The released embedding belongs to a public donor and should not be interpreted as an estimate of the source speaker; source information can instead leak through the source-dependent donor-selection probabilities, which are precisely what Theorem~\ref{thm:thm1} bounds. The donor pool must, however, remain source-independent. Changing the demographic composition of a fixed pool changes the distribution of released surrogate identities but not the DP guarantee, whereas restricting the candidate pool based on the private source (for example, selecting only same-gender donors) can make the output support source-dependent and would invalidate the stated guarantee unless that selection step were itself privatized.

\subsection{Utility}

\begin{figure}[!ht]
    \centering
    \includegraphics[width=0.8\linewidth]{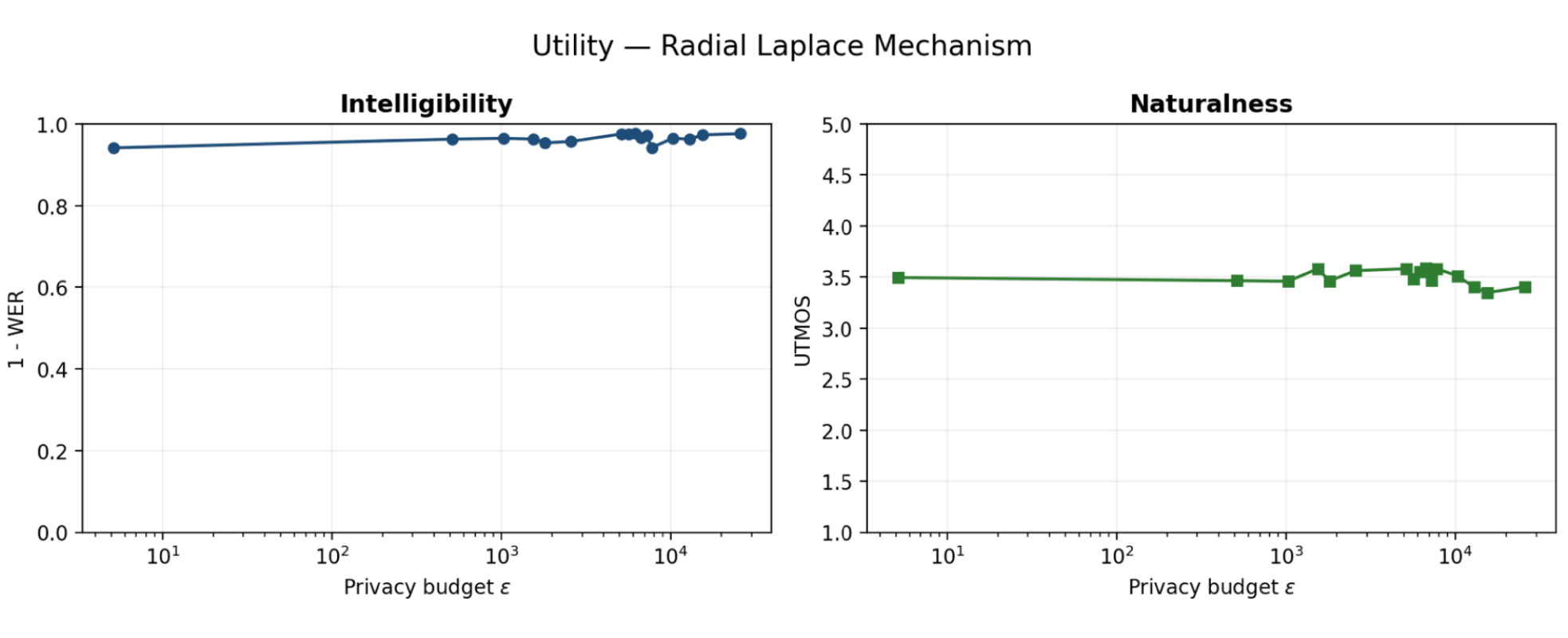}
    \caption{Utility under the radial Laplace mechanism. Word error rate and speech naturalness (MOS) remain largely stable across privacy budgets.}
    \label{fig:WER}
\end{figure}

\begin{figure}[!ht]
    \centering
    \includegraphics[width=0.45\linewidth]{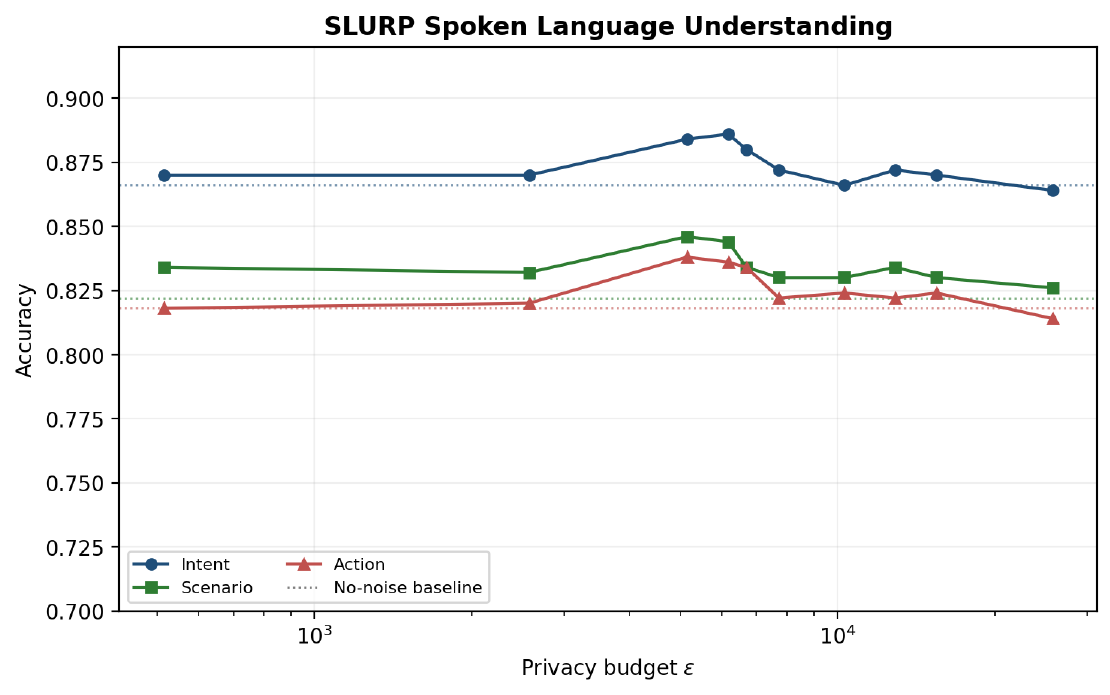}
    \caption{Spoken language understanding on SLURP under the radial Laplace mechanism. Intent, scenario, and action classification accuracy remain stable across privacy budgets.}
    \label{fig:slurp}
\end{figure}

\begin{figure}[!ht]
    \centering
    \includegraphics[width=0.8\linewidth]{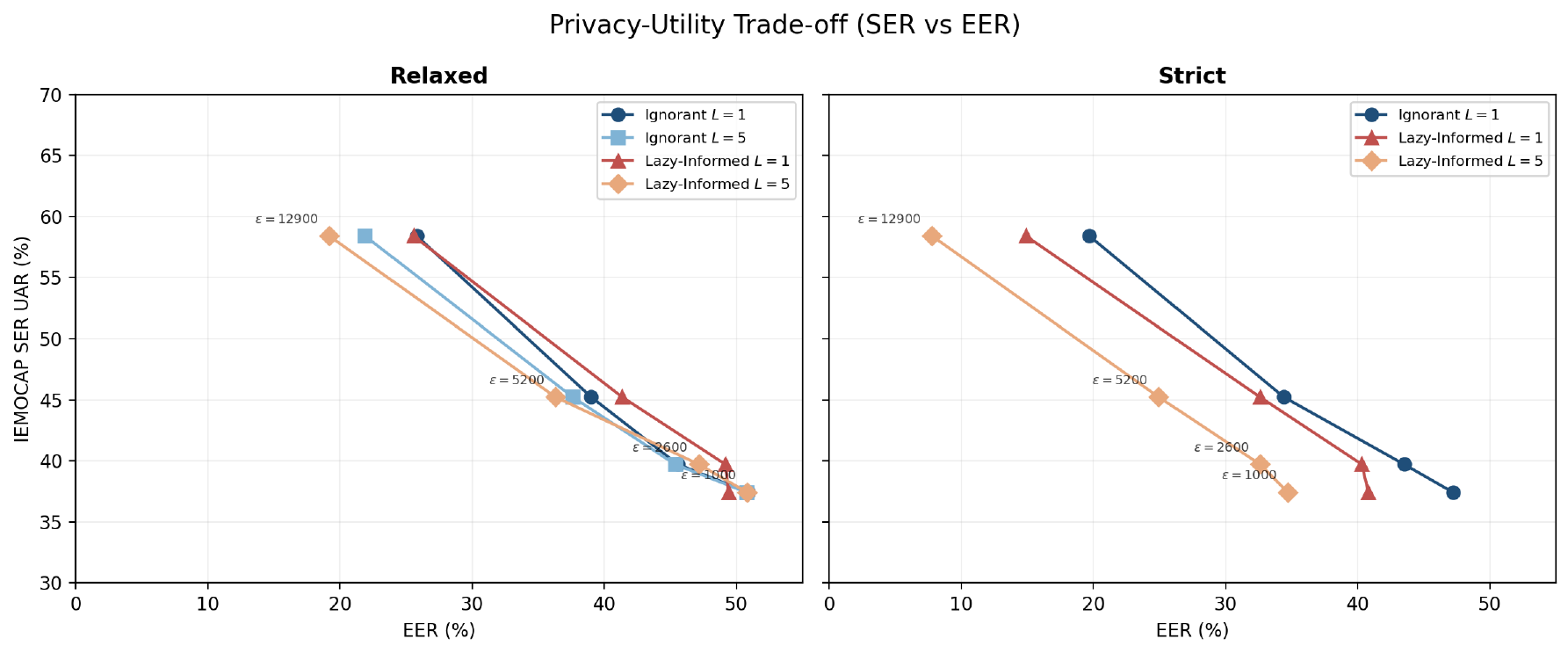}
    \caption{Privacy–utility trade-off between speaker privacy and emotion recognition performance.}
    \label{fig:ser_uar_vs_eer}

\end{figure}

\begin{figure}[!ht]
    \centering
    \includegraphics[width=0.45\linewidth]{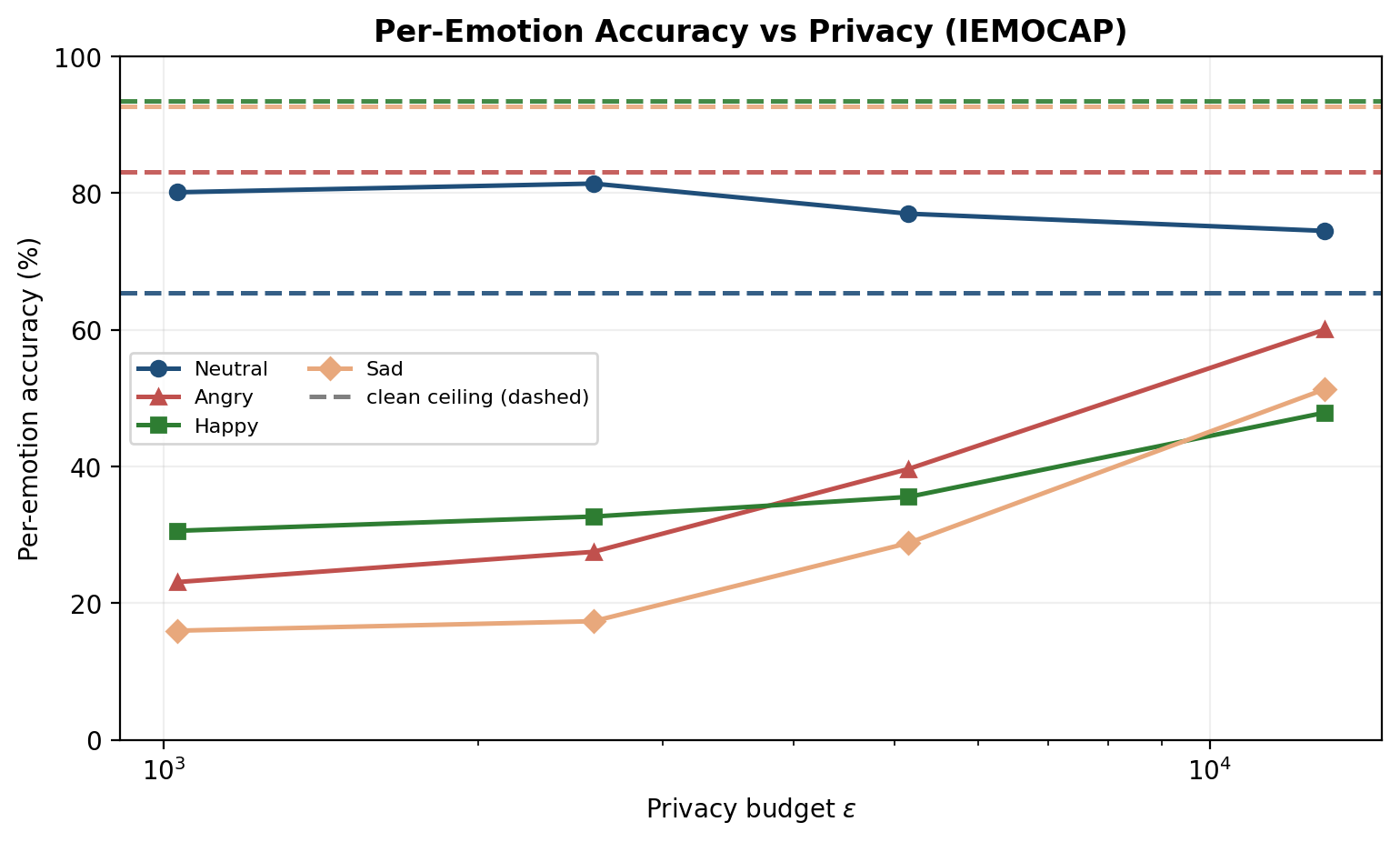}
    \caption{Per-emotion accuracy versus privacy budget. Increasing noise progressively neutralizes the speech signal.}
    \label{fig:ser_per_emotion}
\end{figure}

As a state-of-the-art synthesis backbone, Qwen3-TTS produces high-quality, human-like speech, and this robustness carries over to anonymization across all mechanisms. For the additive mechanisms, WER remains between $0.02$ and $0.06$ across the entire $\varepsilon$ range, indicating that linguistic content is preserved despite the injected noise. Naturalness is likewise stable: the original speech scores approximately $3.3$ on UTMOS, while anonymized speech stays within $3.3$--$3.6$ (Figure~\ref{fig:WER}). Pool selection behaves comparably, with UTMOS holding steady around $3.3$ and WER in the $0.04$--$0.08$ range. Spoken language understanding on SLURP is similarly unaffected across privacy budgets for all mechanisms, with intent, scenario, and action accuracy remaining stable (Figure~\ref{fig:slurp}). Together, these results suggest that anonymization largely preserves both the semantic content and perceptual quality of the generated speech. Emotion recognition is the only downstream task that exhibits a substantial privacy--utility trade-off. For the radial Laplace mechanism, Figure~\ref{fig:ser_uar_vs_eer} shows that stronger privacy systematically reduces UAR. The per-emotion analysis in Figure~\ref{fig:ser_per_emotion} provides further insight into this degradation: recognition accuracy for \emph{neutral} speech improves, whereas accuracy for all expressive emotions declines substantially. This suggests that the mechanism progressively \emph{neutralizes} the speech signal, with the same perturbations that suppress speaker-identifying characteristics also attenuating the prosodic cues required to distinguish emotional states.

\subsection{Source--Surrogate Similarity: Why Not Pick from the Pool at Random?}
\label{app:similarity}

Random donor selection spends no privacy budget, but it discards all information about the source voice. Pool selection contains it as the $\varepsilon \to 0$ endpoint: as $\lambda \to 0$, Algorithm~\ref{alg:pool} samples donors uniformly. To quantify what a nonzero budget buys, we measure how close the selected surrogate is to the source. We expand the donor pool to 100K candidates by interpolation, taking convex combinations of ParaSpeechCaps donor embeddings. All candidates are derived from public donors only, so the pool remains independent of the protected speakers and the guarantee of Theorem~\ref{thm:thm1} is unchanged. For each source utterance, we run Algorithm~\ref{alg:pool} and record the cosine similarity between the source embedding and the selected candidate, averaged over the dataset (Table~\ref{tab:similarity}). Under uniform selection the average similarity is $0.004$, so the surrogate is unrelated to the source. Similarity rises steadily with the budget, reaching $0.665$ at $\varepsilon = 50$ and $0.810$ at $\varepsilon = 500$, against a nearest-donor ceiling of $0.824$. Most of this similarity is already obtained at $\varepsilon = 50$, which is within the range where the mechanism retains high empirical privacy on our main pool (Section~\ref{sec:results}).

\begin{table}[h]
\centering
\small
\begin{tabular}{lc}
\hline
Selection & $\cos(\text{source}, \text{surrogate})$ \\
\hline
Uniform ($\varepsilon \to 0$) & 0.004 \\
$\varepsilon = 10$ & 0.216 \\
$\varepsilon = 50$ & 0.665 \\
$\varepsilon = 500$ & 0.810 \\
Nearest donor ($\varepsilon \to \infty$) & 0.824 \\
\hline
\end{tabular}
\caption{Average cosine similarity between each source embedding and its
selected surrogate, using a pool of 100K candidates. Uniform selection
corresponds to random donor assignment.}
\label{tab:similarity}

\end{table}

\label{app:results}
\label{app:results-utility}

\end{document}